\documentclass[a4paper,11pt]{article}
\usepackage{jcappub} % for details on the use of the package, please see the JINST-author-manual
\usepackage{lineno}
\usepackage{graphicx}
\usepackage{subcaption}
\usepackage{makecell}
\usepackage{multirow}
\usepackage{siunitx}
\usepackage{ulem}
\arxivnumber{}

\title{\boldmath Scattering of electromagnetic field in quasi-topological gravity}

\author{Juan Pablo Arbelaez}
\affiliation{Centro de Matemática, Computação e Cognição (CMCC), Universidade Federal do ABC (UFABC),
Rua Abolição, CEP: 09210-180, Santo André, SP, Brazil}

\emailAdd{juan.arbelaez@ufabc.edu.br}

\abstract{
We investigate the absorption cross sections of electromagnetic perturbations propagating on a four-dimensional brane in the background of higher-dimensional regular black holes arising in quasi-topological gravity. Employing a WKB-based approach for the computation of greybody factors, we analyze the impact of higher-curvature corrections and spacetime dimensionality on the scattering properties of the system. We show that regularity effects lead to systematic deviations from the singular Tangherlini solution, manifested in shifts of the absorption spectrum and modifications of the effective photon sphere radius. Increasing the number of spacetime dimensions suppresses these deviations, driving the system toward the classical limit and reducing the number of multipoles required for convergence. In the high-frequency regime, the absorption cross section approaches the geometric-optics limit, while at low frequencies it is strongly suppressed due to diminished transmission probabilities. Our results demonstrate that the interplay between regularity and higher-curvature effects leaves distinct imprints on the absorption characteristics, providing a sensitive probe of the underlying gravitational theory.}

\begin{document}
\maketitle
\flushbottom

\section{Introduction}
\label{sec:introduction}

Black holes are fundamental objects in gravitational physics, providing a unique arena for probing both classical and quantum aspects of gravity. In particular, the interaction of black holes with external fields, such as scalar, electromagnetic, and Dirac perturbations, plays a crucial role in understanding observable phenomena including Hawking radiation \cite{Hawking:1975vcx}, quasinormal ringing \cite{Nollert:1999ji, Kokkotas:1999bd, Bolokhov:2025rng, Konoplya:2011qq, Berti:2009kk}, and absorption processes. Among these, greybody factors and absorption cross sections encode essential information about the propagation of fields in curved spacetime and serve as sensitive probes of the underlying geometry \cite{Page:1976df,Page:1976ki,Kanti:2002nr}.

In recent years, considerable attention has been devoted to the study of black holes in higher-curvature theories of gravity, motivated both by effective field theory considerations and by the quest for a consistent ultraviolet completion of General Relativity. In this context, quasi-topological gravity provides a particularly attractive framework, allowing for non-trivial higher-curvature corrections while preserving second-order field equations for spherically symmetric configurations \cite{Bueno:2024dgm,Frolov:2024hhe,Bueno:2024eig,Bueno:2025tli,Konoplya:2024kih,Bueno:2026dln, Borissova:2026krh, Borissova:2026wmn, Borissova:2026klg}. Remarkably, such theories admit static, asymptotically flat black hole solutions whose metric functions are determined algebraically, leading to a rich variety of geometries characterized by additional coupling parameters.

An important development within this class of theories is the construction of regular black hole solutions obtained purely from gravitational dynamics, without invoking exotic matter sources. These configurations eliminate curvature singularities and instead exhibit a smooth behavior of curvature invariants throughout spacetime. As a consequence, they provide a natural setting for investigating how deviations from singular geometries affect dynamical and radiative properties of black holes.

The propagation of test fields in such backgrounds reveals several distinctive features. In particular, the effective potential governing perturbations is modified by higher-curvature corrections, which in turn affects the transmission probabilities, greybody factors, and absorption cross sections \cite{Konoplya:2024hfg,Konoplya:2025uta,Arbelaez:2025gwj,Arbelaez:2026eaz,Dubinsky:2026wcv}. At low frequencies, the absorption is typically suppressed due to the dominance of long-wavelength modes, while in the high-frequency regime the cross section approaches the geometric-optics limit determined by unstable photon orbits. Therefore, any modification of the photon sphere structure or of the effective potential barrier is directly reflected in observable scattering characteristics.
%:contentReference[oaicite:0]{index=0}

Furthermore, higher-dimensional setups and brane-world scenarios provide an additional layer of phenomenology, where fields propagate effectively on a lower-dimensional hypersurface while sensing the bulk geometry \cite{Kanti:2004nr}. In such cases, the dimensionality of spacetime and the strength of higher-curvature couplings play a crucial role in shaping the effective scattering barrier and, consequently, the propagation and absorption properties of the fields \cite{Eiroa:2004gh,Kanti:2005ja,Harris:2005jx,Duffy:2005ns,Kanti:2006ua}.

In this work, we investigate the absorption cross sections of test electromagnetic fields propagating on the brane in the background of higher-dimensional regular black holes arising in quasi-topological gravity. Our aim is to quantify the deviations from the corresponding singular solutions and to elucidate the role of higher-curvature corrections and spacetime dimensionality in the scattering process. Particular attention is devoted to the behavior of greybody factors, the convergence of multipolar contributions, and the approach to the geometric-optics limit. We show that the interplay between regularity and higher-curvature effects leads to characteristic modifications of the absorption spectrum, providing a potential observational window into the underlying gravitational theory.

This paper is organized as follows: In Section~\ref{sec:electromag}, we present the basic equations of quasi-topological gravity together with the formalism describing electromagnetic test fields propagating on the brane. In Section~\ref{sec:calculation}, we compute the greybody factors required to determine the absorption cross sections. Section~\ref{sec:absorption} contains the main results of this work, where the absorption properties of higher-dimensional regular black holes are compared with their singular Tangherlini counterparts. In this section, we also analyze the displacement of the photon sphere induced by regularization effects. Finally, in Section~\ref{sec:conclusions}, we summarize the main results and discuss the principal physical implications of our findings.

\section{Electromagnetic perturbations on the brane}
\label{sec:electromag}
Following Ref.~\cite{Bueno:2019ycr}, we consider a higher-curvature gravitational theory whose dynamics are governed by the action
\begin{equation}
    I_{QT} = \frac{1}{16\pi G}\int{d^D x\sqrt{|g|}\left[R+\sum^{n_{max}}_{n=2}{\alpha_n \mathcal{Z}_n}\right]},
    \label{eq:action}
\end{equation}
where \(G\) is the \(D\)-dimensional Newton constant and the quantities \(\mathcal{Z}_n\) denote quasi-topological curvature invariants of order \(n\).
The static, spherically symmetric solutions of this theory can be written in the form \cite{Bueno:2024dgm}:
\begin{equation}
    ds^2 = -N(r)^2 f(r)dt^2 + \frac{dr^2}{f(r)} + r^2d\Omega^2_{D-2},
    \label{eq:metric}
\end{equation}
where \(\Omega^2_{D-2}\) denotes the line element of the unit 
\((D-2)\)-sphere, and, without loss of generality, we choose
\begin{equation}\label{eq:f(r)}
    N(r)=1,\qquad f(r)=1-r^2\psi(r).
\end{equation}
From the equations of motion it follows that the function \(\psi(r)\) satisfies the algebraic equation
\begin{equation}\label{eq:hpsi}
   h(\psi) \equiv \psi + \sum_{n=2}^{n_{max}}{\alpha_{n}\psi^{n}},\quad h(\psi)\equiv\frac{\mu}{r^{D-1}},
\end{equation}
where \(\mu\) is an integration constant that is directly related to the ADM mass of the black hole.

Although the background geometry originates from a higher-dimensional bulk spacetime, the electromagnetic field is assumed to propagate only on the induced four-dimensional brane metric. In this effective description, the angular sector remains that of a two-sphere, and the decomposition of the Maxwell field proceeds in terms of the standard four-dimensional vector spherical harmonics. Consequently, the dimensional dependence enters exclusively through the metric function $f(r)$ inherited from the bulk geometry,
\begin{equation}
    ds^2 = -f(r)dt^2 + \frac{dr^2}{f(r)} + r^2(d\theta^2+\sin^2\theta d\phi^2),
    \label{eq:metric4D}
\end{equation}

We consider electromagnetic perturbations described by a massless spin-1 field propagating on the effective four-dimensional brane geometry governed by the Maxwell equations,
\begin{equation}
    \label{eq:maxwell}
    \frac{1}{\sqrt{-g}} \partial_\mu \Big( \sqrt{-g}F_{\rho\sigma} g^{\rho\nu} g^{\sigma\mu} \Big) = 0,
\end{equation}
where \(F_{\mu\nu} = \partial_\mu A_\nu - \partial_\nu A_\mu\) is the field tensor derived from the vector potential \(A_\mu\). Following Ref.~\cite{Crispino:2000jx}, we adopt Feynman's gauge, which allows for a separation of variables. After decomposing the vector potential into angular modes, the perturbation equations reduce to a one-dimensional Schrödinger-like wave equation,
\begin{equation}\label{eq:wavelike}
    \frac{d^2 \Psi(r_*)}{dr_*^2} + \left[ \Omega^2 - V(r) \right] \Psi(r_*) = 0,
\end{equation}
where \(\Omega\) is the real frequency of the perturbation and \(r_*\) is the tortoise coordinate, defined by
\begin{equation}
\frac{dr_*}{dr} \equiv \frac{1}{f(r)},
\qquad
r_* \to
\begin{cases}
-\infty, & r \to r_0, \\[4pt]
+\infty, & r \to \infty.
\end{cases}
\end{equation}
Here, \(r_0\) denotes the event horizon radius, and the effective potential is
\begin{equation}
    V(r) = f(r)\frac{\ell(\ell+1)}{r^2},
\end{equation}
where \(\ell=1,2,\dots\) is the multipole number.

\begin{figure}[t]
\centering
\begin{subfigure}{0.495\textwidth}
    \centering
    \includegraphics[width=\linewidth]{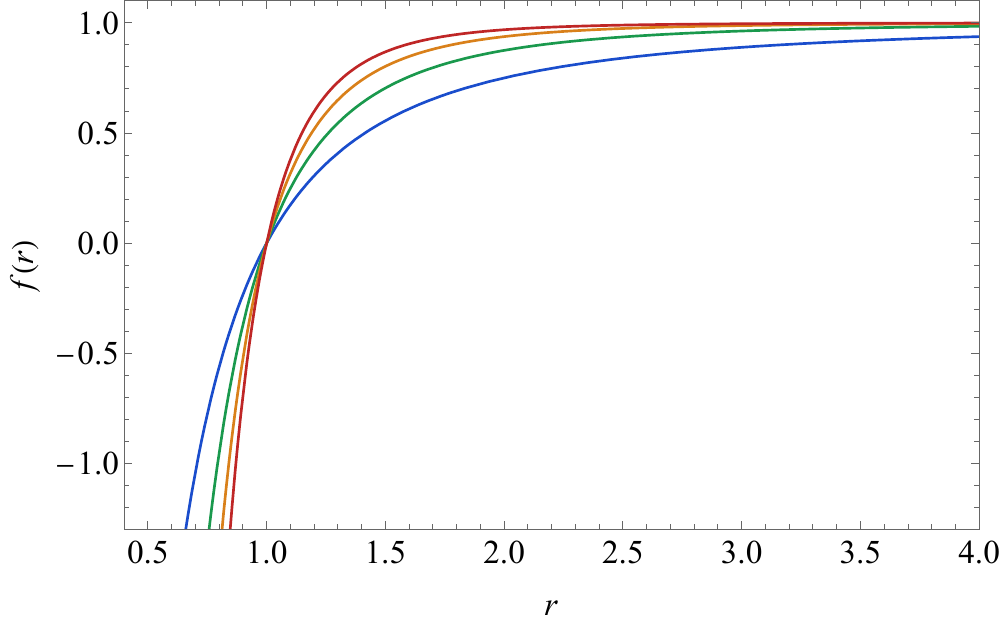}
\end{subfigure}
\hfill
\begin{subfigure}{0.495\textwidth}
    \centering
    \includegraphics[width=\linewidth]{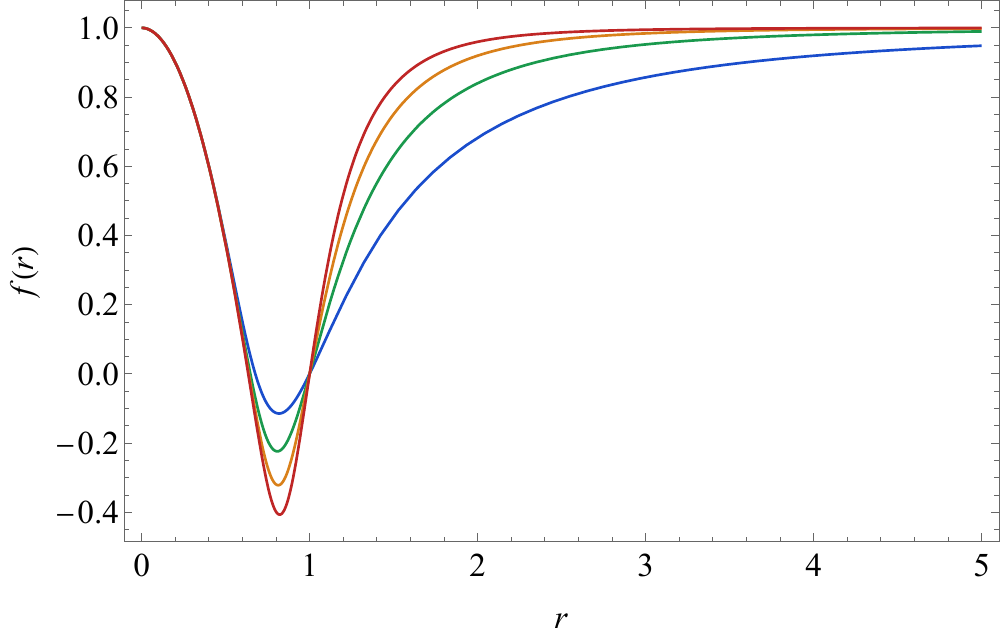}
\end{subfigure}
\caption{Metric functions for the singular Tangherlini solution (left panel) and for the regular configuration $(f)$ (right panel), shown for spacetime dimensions $D=5,6,7,8$. The blue, green, orange, and red curves correspond to $D=5$, $D=6$, $D=7$, and $D=8$, respectively. For the regular solution, the coupling parameter is fixed to $\alpha=0.4$. In both panels, the event-horizon radius is normalized to $r_{0}=1$ in order to work with dimensionless quantities. The figure illustrates how the metric profile changes with increasing dimensionality and how regularization modifies the near-horizon geometry relative to the Tangherlini case.}
\label{fig:veff}
\end{figure}

The metric functions displayed in Fig.~\ref{fig:veff} highlight one of the main geometric differences introduced by regularization, namely the appearance of an inner horizon $r_{i}$ in addition to the event horizon $r_{0}$. Unlike the singular Tangherlini geometry, regular black holes exhibit a modified near-horizon structure characterized by a finite core and a multi-horizon configuration. The most distinctive difference arises in the small-$r$ regime, where regular solutions satisfy
\begin{equation}
 f(r)=1+\mathcal{O}\left(r^2\right)   
\end{equation}
indicating the absence of a central singularity and the recovery of a locally regular core. The presence of the additional horizon changes the radial profile of the metric function in the interior region, producing a richer causal structure and altering the transition between the near-horizon and asymptotic regimes.

\begin{table}[t]
\centering
\small
\setlength{\tabcolsep}{4pt}
\renewcommand{\arraystretch}{1.25}
\begin{tabular*}{\textwidth}{@{\extracolsep{\fill}} c c c c}
\hline
Label 
& $f(r)$ 
& \makecell[c]{$\mu$ parameter in\\horizon units} 
& \makecell[c]{Constraint \\ in $\alpha$} \\
\hline

$(a)$ 
& $\displaystyle 1-\frac{\mu r^2}{r^{D-1}+\alpha \mu}$ 
& $\displaystyle \frac{r_0^{D-1}}{r_0^2-\alpha}$ 
& $\displaystyle 0 \le \frac{\alpha}{r_0^2} \le \frac{D-3}{D-1}$ \\

$(b)$ 
& $\displaystyle 1-\frac{\mu r^2}{\sqrt{r^{2(D-1)}+\alpha^2 \mu^2}}$ 
& $\displaystyle \frac{r_0^{D-1}}{\sqrt{r_0^4-\alpha^2}}$ 
& $\displaystyle 0 \le \frac{\alpha}{r_0^2} \le \sqrt{\frac{D-3}{D-1}}$ \\

$(c)$ 
& $\displaystyle 1-\frac{r^2}{\alpha}\left(1-e^{-\alpha \mu/r^{D-1}}\right)$ 
& $\displaystyle \frac{r_0^{D-1}}{\alpha}\ln\!\left(\frac{r_0^2}{r_0^2-\alpha}\right)$ 
& $\displaystyle 0 \le \frac{\alpha}{r_0^2}\leq C(D)<1$ \\

$(d)$ 
& $\displaystyle 1-\frac{2 \mu r^2}{r^{D-1}+\sqrt{r^{2(D-1)}+4 \alpha^2 \mu^2}}$ 
& $\displaystyle \frac{r_0^{D+1}}{r_0^4-\alpha^2}$ 
& $\displaystyle 0 \le \frac{\alpha}{r_0^2} \le \sqrt{\frac{D-3}{D+1}}$ \\

$(e)$ 
& $\displaystyle 1-\frac{2 \mu r^2}{r^{D-1}+2 \alpha \mu + \sqrt{r^{2(D-1)}+4 \mu \alpha r^{D-1}}}$ 
& $\displaystyle \frac{r_0^{D+1}}{(r_0^2-\alpha)^2}$ 
& $\displaystyle 0 \le \frac{\alpha}{r_0^2} \le \frac{D-3}{D+1}$ \\

$(f)$ 
& $\displaystyle 1-\frac{2 \mu r^2}{\mu \alpha + \sqrt{4 r^{2(D-1)}+\mu^2 \alpha^2}}$ 
& $\displaystyle \frac{r_0^{D-2}}{\sqrt{r_0^2-\alpha}}$ 
& $\displaystyle 0 \le \frac{\alpha}{r_0^2} \le \frac{D-3}{D-2}$ \\

\hline
\end{tabular*}
\caption{Summary of the considered regular black hole models. Configurations $(a)$–$(e)$ were proposed in \cite{Bueno:2024dgm}, while configuration $(f)$ was introduced in \cite{Arbelaez:2025gwj}. The constraint on $\alpha$ for configuration $(c)$, denoted $C(D)$, does not admit a closed-form expression. Its value increases monotonically with $D$ and approaches unity asymptotically, $\lim_{D\to\infty}C(D)=1$. In all cases, the horizon radius $r_{0}$ is defined by the regular horizon condition $f(r_{0})=0$.}
\label{tab:solutions}
\end{table}

\section{Calculation of greybody factors}
\label{sec:calculation}

\begin{figure}[t]
\centering
\begin{subfigure}{0.495\textwidth}
    \centering
    \includegraphics[width=\linewidth]{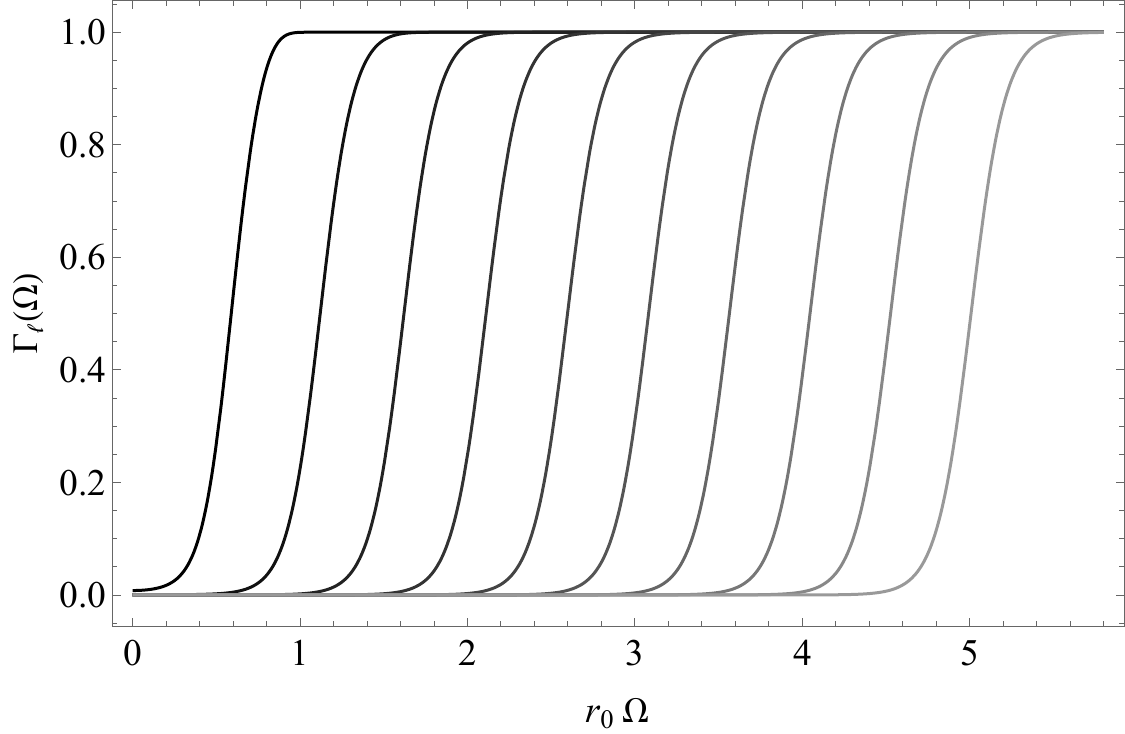}
\end{subfigure}
\hfill
\begin{subfigure}{0.495\textwidth}
    \centering
    \includegraphics[width=\linewidth]{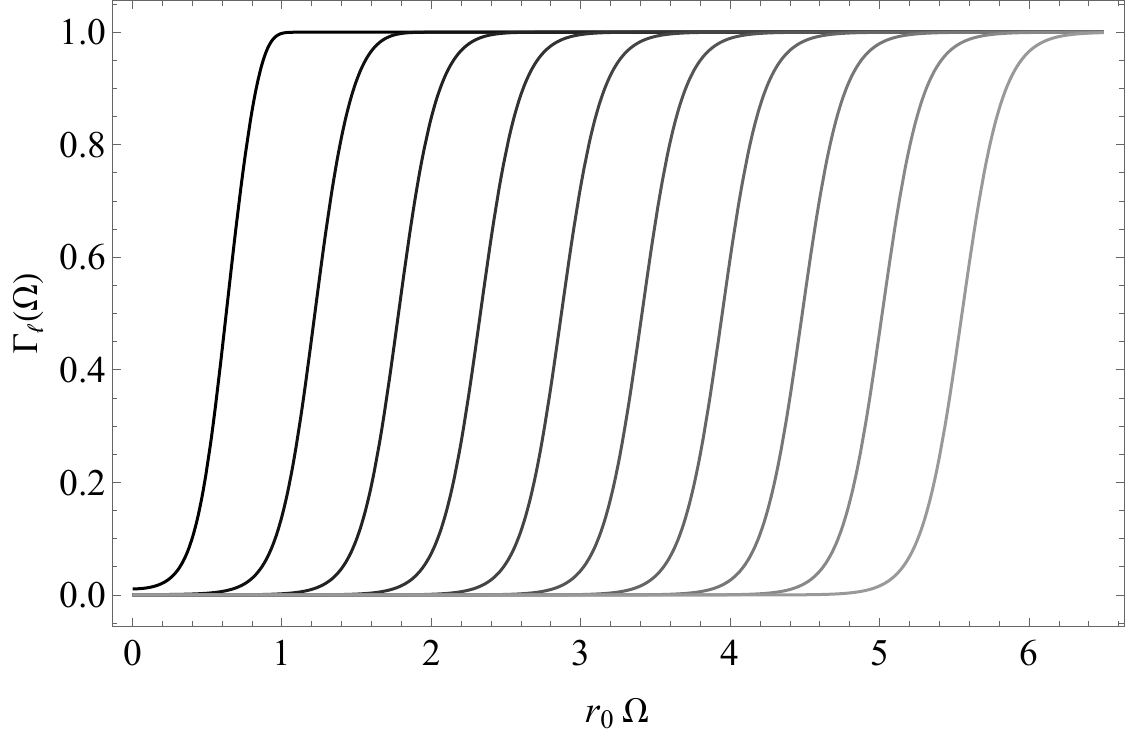}
\end{subfigure}
\caption{Greybody factors for different multipole numbers $\ell$, displayed from darker to lighter tones, corresponding to $\ell=1$ through $\ell=10$, for the configuration $(f)$ with $D=5$ with $\alpha=0.2r_{0}^{2}$ (left panel) and $D=6$ with $\alpha=0.4r_{0}^{2}$ (right panel).}
\label{fig:greybody}
\end{figure}

We consider the scattering problem associated with Eq.~(\ref{eq:wavelike}), subject to the boundary conditions
\begin{subequations}
\begin{align}
    \Psi &= e^{-i\Omega r_{*}} + R_{\ell}\, e^{i\Omega r_{*}}, && r_{*} \to +\infty, \label{eq:psi_infinity} \\
    \Psi &= T_{\ell}\, e^{-i\Omega r_{*}}, && r_{*} \to -\infty. \label{eq:psi_horizon}
\end{align}
\end{subequations}
The greybody factors are defined by
\begin{equation}\label{greybody}
    \Gamma_{\ell}(\Omega)\equiv |T_{\ell}(\Omega)|^2 = 1 - |R_{\ell}(\Omega)|^{2},
\end{equation}
where \(T_{\ell}\) and \(R_{\ell}\) are the transmission and reflection amplitudes, respectively, and their squared moduli give the corresponding transmission and reflection probabilities. These quantities describe the likelihood that an incident wave penetrates the effective potential barrier, thereby connecting the field’s underlying scattering behavior with observable macroscopic quantities such as absorption and emission rates.

Following the general approach of Ref.~\cite{Iyer:1986np}, we employ the WKB approximation formulated as an asymptotic expansion around the eikonal limit \((\ell \to \infty)\), and apply it to the wave-like equation (\ref{eq:wavelike}). Within this framework, the greybody factors can be expressed in terms of the WKB phase as
\begin{equation}
   \Gamma_{\ell}(\Omega) = \frac{1}{1+e^{2i\pi\mathcal{K}}},
\end{equation}
where \(\mathcal{K}\) is a function of the real frequency \(\Omega\). This quantity encodes the local properties of the effective potential, being determined by its value and derivatives evaluated near the maximum. The explicit form of \(\mathcal{K}\) at various orders of the WKB expansion can be found in Refs.~\cite{Iyer:1986np,Konoplya:2003ii,Matyjasek:2017psv}. In principle, the method can be systematically extended to arbitrarily high orders, as demonstrated in Refs.~\cite{Matyjasek:2019eeu,Hatsuda:2019eoj}. In recent years, the WKB approach has been widely applied to the computation of greybody factors for a broad class of compact objects, including black holes and wormholes (see, e.g., Refs.~\cite{Malik:2025dxn,Konoplya:2023ppx,Lutfuoglu:2025ldc,Bolokhov:2024otn, Dubinsky:2024nzo,Konoplya:2019ppy,Lutfuoglu:2025blw,Dubinsky:2025nxv, Lutfuoglu:2025eik,Skvortsova:2024msa,Konoplya:2021ube,Malik:2025erb, Dubinsky:2024vbn,Lutfuoglu:2025hjy,Bolokhov:2025lnt,Konoplya:2010vz, Lutfuoglu:2025kqp,Malik:2024cgb,Lutfuoglu:2025ohb,Konoplya:2023moy, Lutfuoglu:2025ljm}), confirming its robustness and versatility in a variety of geometrical settings.

In the present work, however, we adopt a recently proposed analytic approximation for the WKB phase \cite{Konoplya:2024lir,Konoplya:2024vuj}, which provides a compact expression for the transmission coefficient in terms of the two dominant quasinormal modes, \(\omega_{0}\) and \(\omega_{1}\).These modes are computed using the WKB method (see Refs.~\cite{Konoplya:2019hlu,Konoplya:2026fqh} for a comprehensive review). The resulting approximation for the greybody factor takes the form
\begin{equation}
    \Gamma_{\ell}(\omega) \approx \left[ 1 + \exp\left( \frac{2\pi\bigl(\omega^{2}-\mathrm{Re}(\omega_{0})^{2}\bigr)}{4\,\mathrm{Re}(\omega_{0})\,\mathrm{Im}(\omega_{0})} \right) \right]^{-1}+\mathcal{O}(\ell^{-1}),
    \label{eq:approx}
\end{equation}
which captures the transition between low and high transmission regimes in a simple analytic form. This approximation is already accurate for relatively low multipole numbers and becomes exact in the eikonal limit (\(\ell\to\infty\)), where the WKB expansion is formally justified.

As pointed out in Ref.~\cite{Arbelaez:2026eaz}, the accuracy of this method is sufficient for the range of parameters explored in the present work. In the calculation of the greybody factors (see Fig.~\ref{fig:greybody}), we evaluate the greybody factors modes using the approximate formula up to order \(\mathcal{O}(\ell^{-3})\).

\section{Absorption cross section}
\label{sec:absorption}
\begin{figure}[t]
\centering
\begin{subfigure}{0.495\textwidth}
    \centering
    \includegraphics[width=\linewidth]{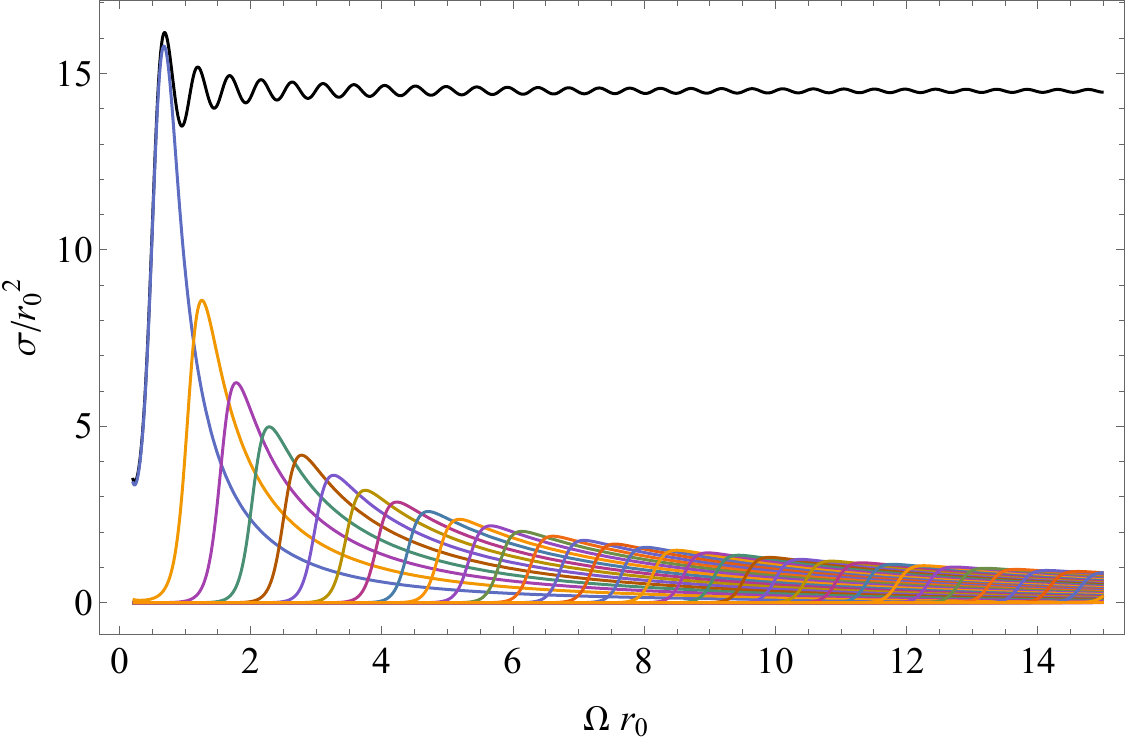}
\end{subfigure}
\hfill
\begin{subfigure}{0.495\textwidth}
    \centering
    \includegraphics[width=\linewidth]{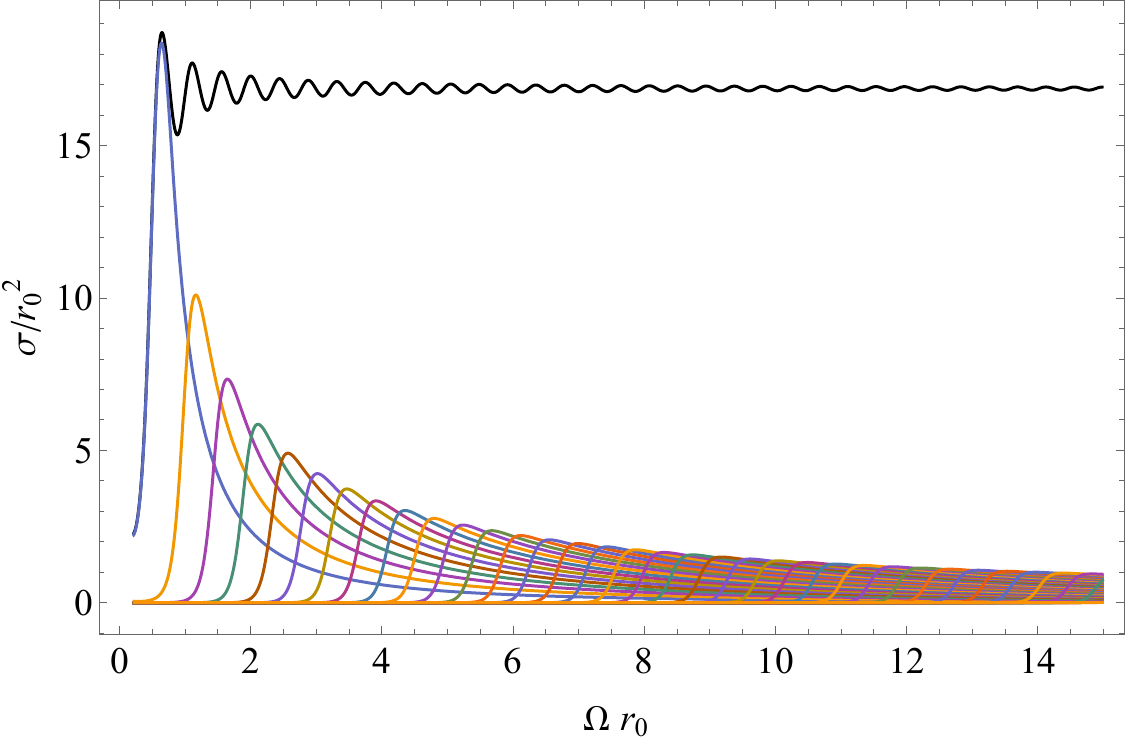}
\end{subfigure}
\caption{Contribution of the first 50 multipoles for configuration $(f)$ with $D=5$. The left panel corresponds to $\alpha/r_{0}^{2}=0.3$, while the right panel corresponds to $\alpha/r_{0}^{2}=0.5$.}
\label{fig:crossD5}
\end{figure}
The absorption cross section $\sigma_{\rm abs}$ provides a direct measure of the interaction between incident electromagnetic radiation and the black hole geometry, encoding how efficiently incoming modes are captured after traversing the effective scattering barrier. Within the partial-wave formalism, the total absorption is obtained by summing the transmission probabilities associated with each angular mode~\textbf{\cite{Crispino:2010fd}},
\begin{equation}
    \sigma_{\rm abs}(\Omega)=\frac{\pi}{\Omega^{2}}
    \sum_{\ell=1}^{\infty}(2\ell+1)\Gamma_{\ell}(\Omega),
\end{equation}
where $\Gamma_{\ell}(\Omega)$ are the greybody factors associated with each multipole number $\ell$ and frequency $\Omega$. 

At low frequencies, only the lowest multipoles contribute significantly, leading to a rapid decrease of the cross section as $\Omega$ decreases. This behavior is further enhanced by the suppression of the greybody factors in this regime. In particular, since $\Gamma_\ell(\Omega)\to 0$ as $\Omega\to 0$ for $\ell\ge1$, the low-frequency contribution to the integrated flux is naturally suppressed. From a physical perspective, this is consistent with the fact that, in the low-frequency regime, the photon wavelength is much larger than the characteristic black-hole scale, namely $\lambda \gg r_{0}$. As a consequence, the incoming radiation couples only weakly to the near-horizon region, and the black hole does not exhibit a finite effective absorption area for very low-energy photons (see Fig.~\ref{fig:crossD5}).

In the high-frequency regime, wave propagation reduces to geometric optics and the black hole behaves as a classical absorber. In this limit, $\sigma_{\rm abs}$ approaches the geometric-optics capture cross section, $\sigma_{\rm geo} = \pi b_c^2,$ where $b_c$ is the critical impact parameter associated with unstable photon orbits. Any modification of the geometry induced by the regularizing parameter or by the spacetime dimension is thus directly reflected in the value of $b_c$ and, consequently, in the limiting value of the absorption cross section. In addition, the approach to $\sigma_{\rm geo}$ is typically accompanied by oscillatory features, which can be interpreted as wave effects around the classical capture threshold and are closely related to the presence of unstable photon orbits. Particles with $b < b_c$ are captured by the black hole, whereas those with $b > b_c$ are scattered to infinity. Therefore, the critical impact parameter $b_c$ defines the apparent shadow radius for an asymptotic observer.

For null geodesics in a spherically symmetric spacetime (\ref{eq:metric4D}), one may restrict the motion to the equatorial plane ($\theta=\pi/2$) without loss of generality. The conserved quantities associated with time translation and rotational symmetries are the energy and angular momentum,
\begin{equation}
    E=-g_{tt}\dot{t}=f(r)\dot{t},\qquad 
    L^{2}=r^{4}\dot{\phi}^{2}.
\end{equation}
The radial motion of photons is governed by
\begin{equation}
    \dot{r}^{2}+V_{\mathrm{eff}}(r)=E^{2},
\end{equation}
where the effective potential is
\begin{equation}
    V_{\mathrm{eff}}(r)=f(r)\frac{L^{2}}{r^{2}}.
\end{equation}
For an asymptotic observer, photon trajectories are characterized by the impact parameter
\begin{equation}
    b=\frac{L}{E}.
\end{equation}
The impact parameter determines whether an incoming photon is scattered back to infinity or captured by the black hole, depending on the relation between $E^2$ and the effective potential barrier. The critical impact parameter $b_c$ is associated with unstable circular null geodesics; these orbits occur at the photon-sphere radius $r_{\rm ph}$, defined by the conditions
\begin{equation}
    \dot r=0, \qquad V'_{\rm eff}(r_{\rm ph})=0,
\end{equation}
together with the instability requirement
\[
V''_{\rm eff}(r_{\rm ph})<0.
\]
Evaluating the radial equation of motion at $r=r_{\rm ph}$, one obtains
\begin{equation}
    b_{c} = \frac{r_{\rm ph}}{\sqrt{f(r_{\rm ph})}}.
\end{equation}

\begin{figure}[t]
\centering
\begin{subfigure}{0.495\textwidth}
    \centering
    \includegraphics[width=\linewidth]{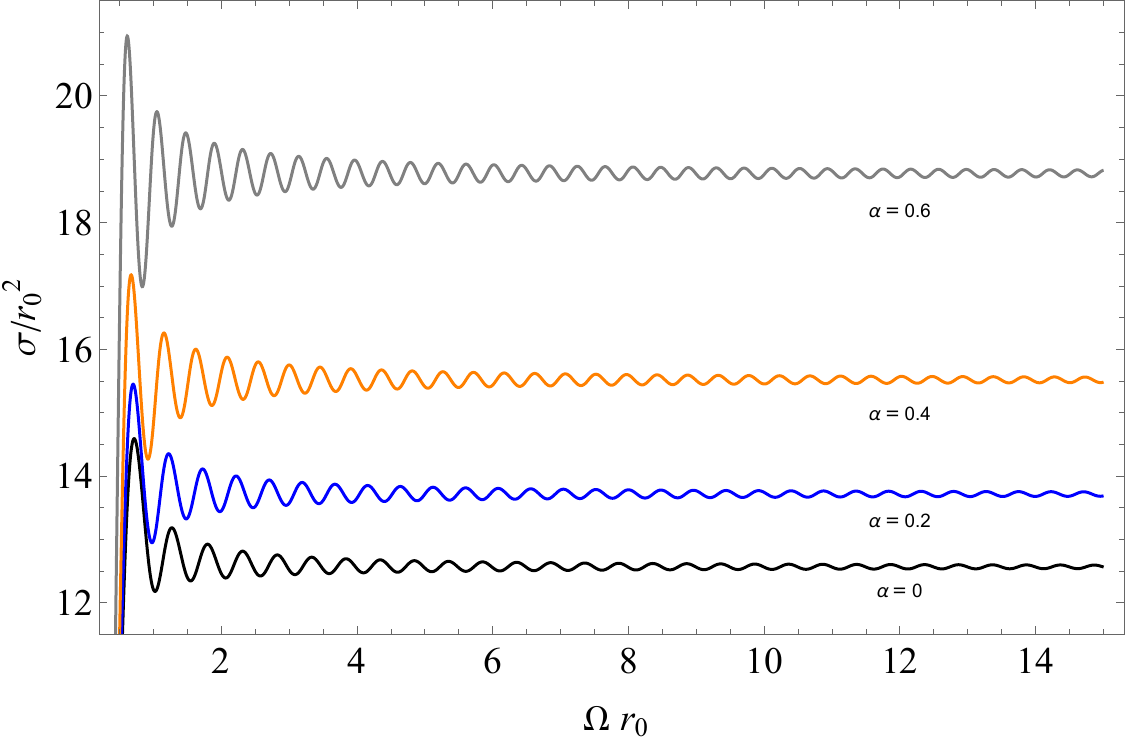}
\end{subfigure}
\hfill
\begin{subfigure}{0.495\textwidth}
    \centering
    \includegraphics[width=\linewidth]{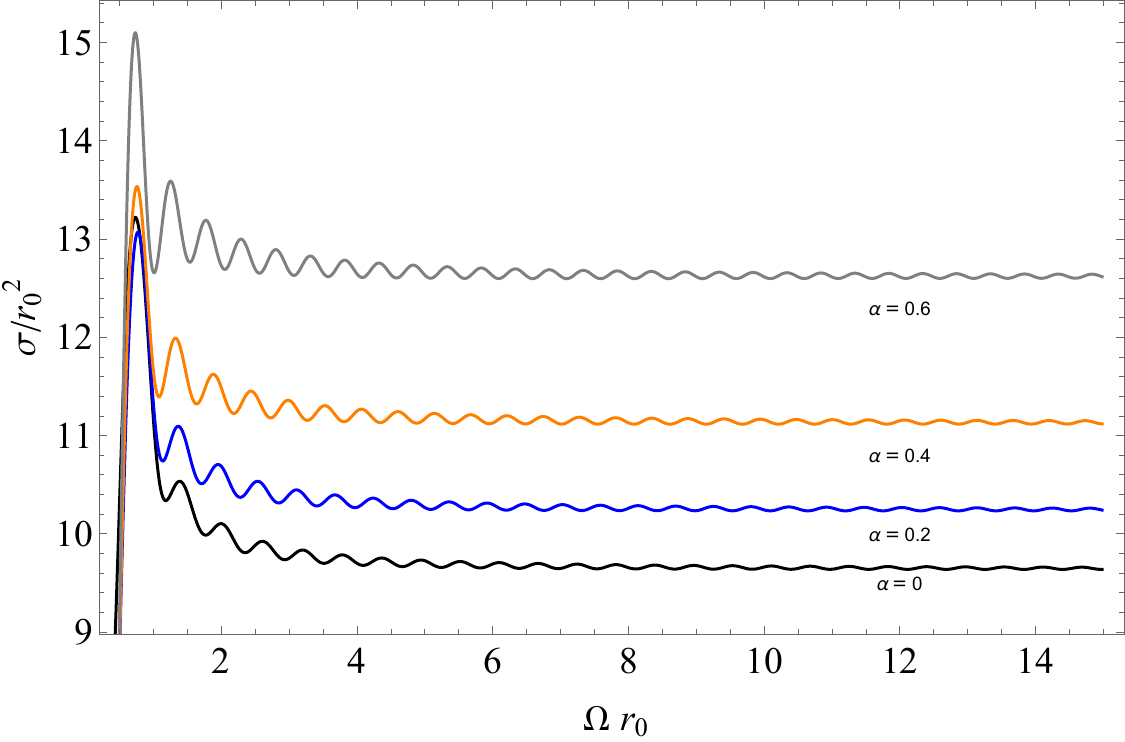}
\end{subfigure}
\caption{Contributions on total cross-section for the first 40 modes for the $(f)$ black hole in $D=5$ (left panel) and $D=6$ (right panel) for some values of $\alpha/r_{0}^{2}$.}
\label{fig:totalcross}
\end{figure}

From Fig.~\ref{fig:totalcross}, one observes a clear shift in the total contributions to the absorption cross section, together with a deviation from the singular Schwarzschild--Tangherlini solution as the ratio $\alpha/r_{0}^{2}$ increases. This deviation persists throughout the full admissible range of the coupling parameter and remains present up to the extremal regime. The same qualitative trend is observed across all spacetime dimensions and for each regular configuration considered. Furthermore, the relative contribution of the lowest multipoles becomes increasingly dominant as the spacetime dimensionality grows. As the extremal regime is approached, an increasing number of multipoles is required in order to achieve convergence to the geometric optics limit.

Meanwhile, Table~\ref{tab:photon_sphere} summarizes the effective absorption cross section in the geometric optic regime. A clear and robust trend emerges: the impact parameter of the circular photon orbit decreases as the spacetime dimensionality increases, indicating an inverse scaling with the number of dimensions. This behavior can be understood from the fact that, in higher dimensions, the gravitational field decays more rapidly with distance, effectively localizing its influence closer to the black hole. As a consequence, unstable photon orbits occur at smaller radii, leading to a reduced photon sphere.

This tendency is already present in the singular Tangherlini solution, which provides a useful baseline for comparison. In this case, the metric function is given by 
$$f(r) = 1 - \left(\frac{r_0}{r}\right)^{D-3},$$
and the photon sphere radius can be obtained analytically. Imposing the condition $V_{\mathrm{eff}}'(r_{\text{ph}})=0$ yields
\begin{equation}
\frac{r_{\text{ph}}}{r_{0}} = \left( \frac{D-1}{2} \right)^{\frac{1}{D-3}},
\label{eq:tangD}
\end{equation}
which decreases monotonically toward unity as $D$ increases. The regular black hole models considered here follow the same qualitative behavior, with deviations controlled by the higher-curvature parameter.

\begin{figure}[t]
\centering
\begin{subfigure}{0.495\textwidth}
    \centering
    \includegraphics[width=\linewidth]{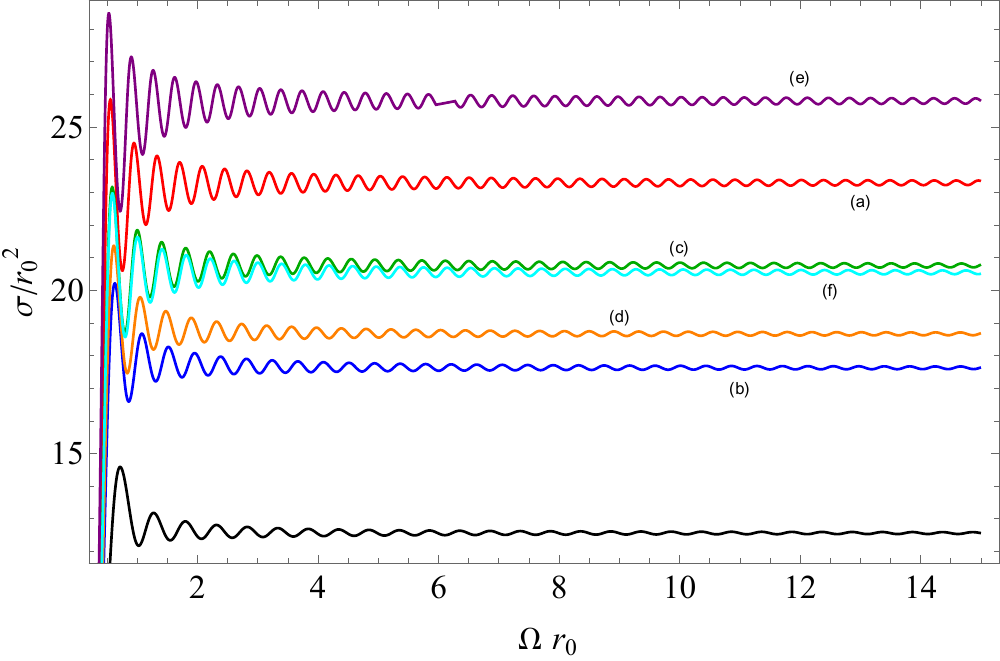}
\end{subfigure}
\hfill
\begin{subfigure}{0.495\textwidth}
    \centering
    \includegraphics[width=\linewidth]{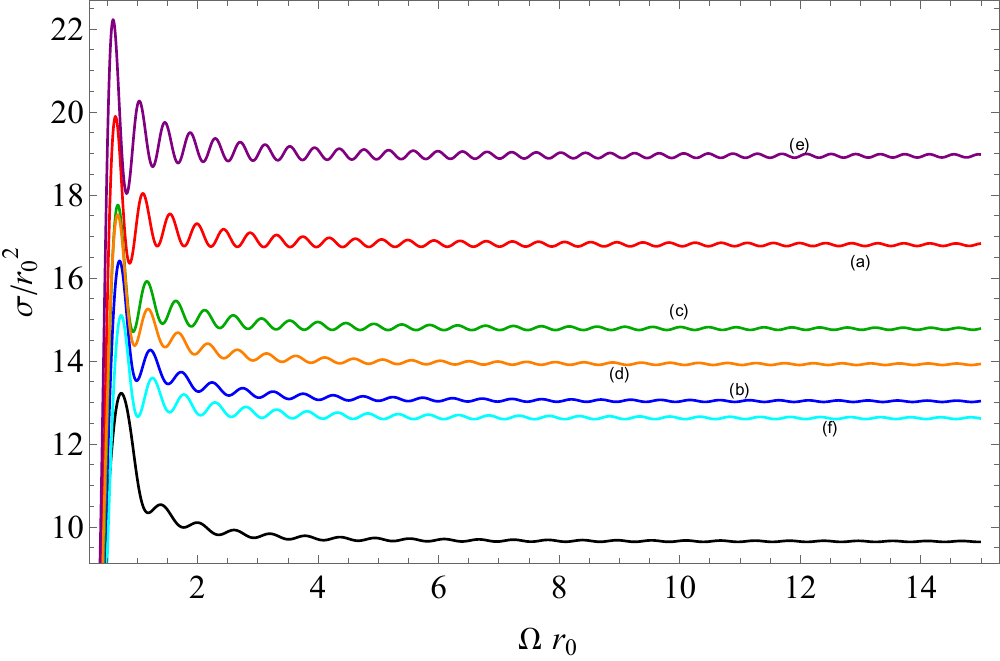}
\end{subfigure}
\caption{First 40 contributions to the total cross section for the configurations presented in Table~\ref{tab:solutions} for $D=5$ (left panel) and $D=6$ (right panel). The lower black line corresponds to the singular Tangherlini solution, red for $(a)$, blue for $(b)$, dark green for $(c)$, orange for $(d)$, purple for $(e)$, and cyan for $(f)$. All configurations are in their extremal regime in terms of the coupling parameter $\alpha/r_{0}^{2}$.}
\label{fig:Allcontributions}
\end{figure}

Complementing the results presented in Fig.~\ref{fig:totalcross}, we computed the total contributions of the first 40 multipoles for extremal configurations characterized by a vanishing Hawking temperature, $T_{H}(r_{0})=0$, and compared them with the singular Tangherlini solution in Fig.~\ref{fig:Allcontributions} for all configurations. Together with the photon sphere radii compiled in Table~\ref{tab:photon_sphere}, a clear trend is observed: the differences with respect to the singular solution decrease as the bulk dimensionality increases. As $D$ grows, the discrepancies between the regular extremal configurations and the singular Tangherlini solution are systematically suppressed. This behavior suggests that $D$ acts as a parameter controlling the localization of photon trajectories projected onto the brane, restricting the range of allowed impact parameters and driving the dynamics of test fields toward the singular limit as $D$ increases. This result is consistent with the behavior encoded in Eq.~(\ref{eq:tangD}), where the photon sphere radius approaches its asymptotic value as the number of dimensions grows. As a consequence, the effective potential projected onto the brane converges toward that of the singular Tangherlini solution, leading to the observed agreement.
\begin{table}[h]
\centering
\renewcommand{\arraystretch}{1.2}
\setlength{\tabcolsep}{8pt}

\begin{tabular}{|c|c|c|c|c|c|}
\hline
\multicolumn{2}{|c|}{\textbf{Model}} & \multicolumn{4}{c|}{$\mathbf{\pi b_{c}^{2}}$} \\
\cline{3-6}
\multicolumn{2}{|c|}{} & $\mathbf{D=5}$ & $\mathbf{D=6}$ & $\mathbf{D=7}$ & $\mathbf{D=8}$ \\
\hline

\multicolumn{2}{|c|}{Tangherlini} & 12.56640 & 9.64477 & 8.16210 & 7.25947 \\
\hline

\multirow{2}{*}{$(a)$} 
& $\alpha=0.2$ & 15.01920 & 10.89260 & 8.95940 & 7.83183 \\
\cline{2-6}
& $\alpha=0.4$ & 19.48270 & 12.92620 & 10.18880 & 8.68538 \\
\hline

\multirow{2}{*}{$(b)$} 
& $\alpha=0.2$ & 12.81000 & 9.77090 & 8.24282 & 7.31725 \\
\cline{2-6}
& $\alpha=0.4$ & 13.65180 & 10.19860 & 8.51401 & 7.51029 \\
\hline

\multirow{2}{*}{$(c)$} 
& $\alpha=0.2$ & 13.68800 & 10.22940 & 8.54008 & 7.53271 \\
\cline{2-6}
& $\alpha=0.4$ & 15.34880 & 11.04930 & 9.05553 & 7.89890 \\
\hline

\multirow{2}{*}{$(d)$} 
& $\alpha=0.2$ & 13.05940 & 9.89895 & 8.32445 & 7.37555 \\
\cline{2-6}
& $\alpha=0.4$ & 14.84950 & 10.78910 & 8.88293 & 7.77063 \\
\hline

\multirow{2}{*}{$(e)$} 
& $\alpha=0.2$ & 18.19090 & 12.36530 & 9.85984 & 8.46180 \\
\cline{2-6}
& $\alpha=0.4$ & - & 17.75110 & 12.87970 & 10.46780 \\
\hline

\multirow{2}{*}{$(f)$} 
& $\alpha=0.2$ & 13.71570 & 10.24320 & 8.54876 & 7.53885 \\
\cline{2-6}
& $\alpha=0.4$ & 15.51890 & 11.12970 & 9.10465 & 7.93311 \\
\hline

\end{tabular}
\caption{Geometric-optics absorption cross section $\sigma_{\rm geo}=\pi b_c^2$ for the Tangherlini solution and the regular black hole configurations considered in this work. Results are shown for several spacetime dimensions with $r_0=1$, illustrating the systematic enhancement of the optical capture area induced by regularization.}
\label{tab:photon_sphere}
\end{table}

Overall, the results presented in this section indicate that the bulk dimensionality $D$ and the higher-curvature coupling $\alpha$ play complementary roles in shaping the absorption properties of regular black holes projected onto the brane. While increasing $D$ progressively suppresses deviations from the singular Tangherlini geometry, the parameter $\alpha$ quantifies the departure from the singular limit by controlling the strength of the regularization effects. In particular, larger values of $\alpha$ produce an outward displacement of the photon sphere, increasing the critical impact parameter and consequently enlarging both the shadow radius and the geometric-optics absorption cross section.

\section{Conclusions}\label{sec:conclusions}
In this work, we have investigated the absorption cross sections for electromagnetic test fields propagating on a 4D brane, where the background geometry is determined by higher-dimensional regular black holes in the bulk. Our analysis focused on quantifying deviations from the singular Tangherlini solution and on assessing the role of the higher-curvature coupling parameter $\alpha$.

It is found that, for all values of $\alpha/r_{0}^{2}$, the shadow radius of regular black holes increases relative to their singular counterparts. This behavior reflects a significant enlargement of the photon capture region through an outward displacement of the photon sphere, driven by modifications of the effective potential induced by the regular geometry. The regularization of the central core alters the internal spacetime structure, shifting unstable null orbits to larger radii through purely geometric mechanisms and enhancing the effective gravitational trapping region. As a consequence, both the shadow radius and the absorption cross section increase, while the greybody spectrum is displaced toward higher frequencies, indicating that more energetic modes are required to efficiently penetrate the modified effective potential barrier. The shadow radius continues to increase up to the extremal configuration determined by the spacetime dimension $D$, at which the inner $(r_{i})$ and event $(r_{0})$ horizons merge into a degenerate horizon.

From a more complete physical perspective, the displacement observed in the brane-projected absorption cross section curves across all frequency regimes can be directly attributed to the behavior of the greybody factors, namely to a deformation and effective shift of the scattering barrier, as previously discussed in \cite{Arbelaez:2026eaz}. In this sense, the enhancement of the classical photon capture region, the modification of the wave-scattering barrier, and the change in the effective size of the black hole should not be interpreted as independent effects. Rather, they arise simultaneously as a consequence of the deformed spacetime geometry induced by the regularization mechanism. The regular core modifies the effective potential governing null geodesics and wave propagation alike, producing a coherent geometric response observable in shadow formation, absorption spectra, and transmission probabilities.

We find that increasing the bulk dimensionality $D$ systematically reduces both the number of multipoles required to achieve convergence of the total absorption cross section and the deviation of the effective photon sphere radius with respect to the singular case. This behavior reflects the fact that, in higher dimensions, the gravitational field becomes more localized near the horizon. Consistently, the effective potential barrier projected onto the brane becomes narrower and higher as $D$ increases, leading to a suppression of the greybody factors, particularly in the low-frequency regime. As a result, long-wavelength modes exhibit a reduced transmission probability, and their contribution to the absorption cross section rapidly decreases. The simultaneous suppression of multipolar contributions and the convergence of the photon sphere radius toward the Tangherlini limit can thus be understood as complementary manifestations of the same underlying mechanism.

Overall, our results indicate that the bulk dimensionality acts as a controlling parameter governing the departure from General Relativity in brane-projected regular black hole geometries. As $D$ increases, the system progressively approaches the singular Tangherlini limit, while the geometric effects induced by regularization become less pronounced. At the same time, the coupling parameter controls an effective enlargement of the photon trapping region, leading to an increase in the optical size of the black hole. These results suggest that shadow observables, greybody factors, and absorption spectra constitute complementary probes of the same underlying geometric deformation induced by higher-dimensional regularization mechanisms, offering a unified framework to characterize departures from singular black-hole geometries.

\acknowledgments
J. P. A. acknowledges support from the Conselho Nacional de Desenvolvimento Científico e Tecnológico (CNPq). The author is also grateful to Alexander Zhidenko for proposing the problem and reading the manuscript.

\bibliographystyle{JHEP}
\bibliography{bibliography.bib}

\end{document}